\documentclass[11pt,twoside]{article}

\usepackage{asp2010}
\resetcounters
\pdfoutput=1
\begin{document}
\title{Faint Submillimeter Galaxies behind the Massive Lensing Cluster A2390}
\author{Chian-Chou Chen$^1$ and Lennox Cowie$^1$
\affil{$^1$Institute for Astronomy, University of Hawaii, 2680 Woodlawn Drive, Honolulu, HI 96822, USA}}

\begin{abstract}
Current studies on Submillimeter Galaxies (SMGs) mostly focus on bright sources with 850~$\mu$m flux greater than 2~mJy, and the results have shown that they are likely high redshift mergers with z $>$ 2 and could be a dominant population on star formation in the early Universe. However, bright SMGs only contributes 20-30\% of the 850~$\mu$m extragalactic background light (EBL), meaning the bulk of the cosmic star formation still hidden by dust and our current understanding is biased. We have started a program to study an unbiased sample of highly-amplified and intrinsically faint SCUBA detected SMGs in the field of massive lensing clusters. Here we report the newly obtained SMA observations at 850~$\mu$m on one of our sample source, A2390-5, behind the massive lensing cluster A2390. We successfully detect the source with a flux of 3.95~mJy. Surprisingly, it does not have any counterpart in any other wavelengths even though there are tentative candidates, which implies a very dusty and high-z nature. With less than 1$\arcsec$ positional accuracy and the adoption of z = 5, we obtain the amplification factor of 12 using current lensing model, which makes A2390-5 a faint SMG with a de-lensed flux of 0.33~mJy. Together with our previous detection on another faint SMG, both of them have no counterpart in other wavelengths and their properties are very different than previously thought from the single-dish data. We emphasize the importance of direct submillimeter high-resolution studies on faint SMGs, which could be the dominant population of the high-z star formation.
\end{abstract}

\section{Introduction}
Blank-field submillimeter surveys with ground-based single-dish telescopes have resolved a rare but cosmologically important population, submillimeter galaxies 
(SMGs; \citealt{Blain:2002p8120}). Detail follow-up studies on both continuum and emission lines have shown that SMGs are at high redshift (2$<$z$<$5), gas rich ($M_{gas} > 10^{10} M_\odot$), relatively short lived ($<$~100 Myr), highly clustering, likely mergers and the progenitors of massive elliptical galaxies in present-day Universe. Despite the rareness, their extreme star formation (SFR $\sim$ 100$-$1000 M$_\odot$ yr$^{-1}$ ) makes them a major contributor to star formation in the early Universe 
(\citealt{Barger:2000p2144}).

However, the blank-field SMGs currently under extensive study are usually bright with 850~$\mu$m fluxes above 2~mJy, and they only contribute 20-30\% of the 850~$\mu$m extragalactic background light (EBL; \citealt{Coppin:2006p9123}), which is the integrated light from sources outside our Milky Way galaxy. Thus bulk of the cosmic star formation is still unresolved and the formation mechanism of the faint SMGs emitting the rest of the 850~$\mu$m EBL naturally becomes a key issue to address some important questions such as are mergers the major contributor to the cosmic stellar assembly in the early Universe. 

Unfortunately, the poor resolution (e.g., $\sim14\arcsec$ FWHM on the JCMT at 850~$\mu$m) prevents us from directly measuring the faint SMGs below the 2~mJy confusion limit. Almost all of our knowledge about faint SMGs comes from observations with single-dish telescopes in the fields of massive lensing clusters. Due to the presence of the intervening cluster mass, the intrinsically faint fluxes of background sources are gravitationally amplified to a detectable level, and the confusion limit is reduced by the expansion of the source plane. Faint SMGs with fluxes between 0.1 and 2~mJy have been detected in this way, and they contribute more than 50\% of the 850~$\mu$m EBL (\citealt{Cowie:2002p2075, Knudsen:2008p3824}). As with the bright SMGs, pinning down their exact location is the first and the most crucial step for any follow-up studies. We have started a program to study an unbiased sample of highly-amplified and intrinsically faint SCUBA detected SMGs in the field of massive lensing clusters. Here we discuss the newly obtained high-resolution Submillimeter Array (SMA) observations at 850~$\mu$m on one of our sample source, A2390-5, behind the massive lensing cluster A2390.

\section{SMA observations on A2390-5}

\begin{figure}
 \begin{center}
    \leavevmode
     \includegraphics[scale=0.15]{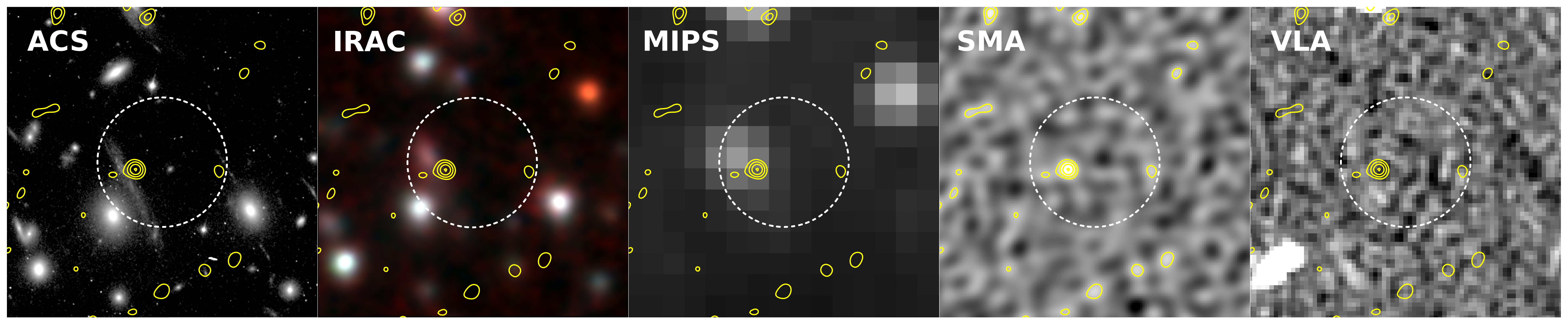}
 	\end{center}
       \caption{Postage stamp images, centered on the SMA phase center position of A2390-5, which is the original SCUBA centroid from \citet{Cowie:2002p2075}. From left to right: ACS f850lp, IRAC false color 3.6-4.5-5.8~$\mu$m, MIPS 24~$\mu$m, SMA 850~$\mu$m and VLA~1.4 GHz imaging with a size of 36$'' \times 36''$. SMA signal is shown in yellow contours with levels at 2, 3, 4, 5 $\sigma$ in each panel. North is up and east is to the left. Dashed-white circles show SCUBA beams with 7\farcs5 in radius.}
     \label{a23905}
\end{figure}

Two tracks of data were taken in 2011, one in May and the other in September. To increase the signal-to-noise ratio we use compact configuration and make final continuum images using both tracks and naturally weighted baselines. We used the data reduction package MIR to calibrate the visibilities and produced the images using the MIRIAD routines. In the end the theoretical noise at the phase center is 0.74 mJy/beam and the FWHM of the synthesized beam is 2$\farcs$2$\times$1$\farcs$8. 

A2390-5 was first detected by \citealt{Cowie:2002p2075} with a 850~$\mu$m flux of 2.64~mJy from a SCUBA survey and the amplification factor is 39 due to its proximity to a cluster member. Melcalfe et al. (2003) also report detections at 7 and 15~$\mu$m using ISOCAM on the arc structure enclosed by SCUBA beam, a very likely counterpart for A2390-5. However our SMA observations reveal a surprising different story.

Figure \ref{a23905} shows the postage stamp images of A2390-5. The SMA signal is shown in yellow contours while color maps in each panel show the data from optical to radio at the same region. North is up and East is to the left. The only possible counterpart of A2390-5 is the north-east component relative to the submillimeter 3.95~mJy (5$\sigma$) detection seen in the IRAC and MIPS map within the dashed-white circle representing the SCUBA beam. However referring to the ACS image it is more likely that it traces part of the optical arc instead of the submillimeter signal from A2390-5, based on the fact that the morphology of the infrared signal is elongated aligned with the arc and the position offset is much greater than the positional error measured from the SMA. Note that among the data from other wavelengths, radio maps are usually the best tracer for the SMGs (\citealt{Condon:1992p6652}) and the submillimeter/radio flux ratio has been used as a proxy to the redshifts (\citealt{Wang:2007p6971, Cowie:2009p6978, Chen:2011p11605}). A higher ratio implies a higher redshifts. For A2390-5, a non-detection in a very deep radio map (1$\sigma$ = 6.5~$\mu$Jy) gives us a redshift $\ge$ 5.8. With less than 1$\arcsec$ positional accuracy and the adoption of z = 5, we obtain the amplification factor of 12 using the current lensing model, which makes A2390-5 a faint SMG with a de-lensed flux of 0.33~mJy, much larger than the previous estimate. Under the assumption of a spectral energy distribution from \citet{Barger:2000p2144} based on an Arp 220 template, the total IR luminosity of A2390-5 is 3.6 $\times$ 10$^{11} L_\odot$, which is a more typical star forming galaxy compared to bright SMGs.

\begin{figure}
 \begin{center}
    \leavevmode
     \includegraphics[scale=0.95]{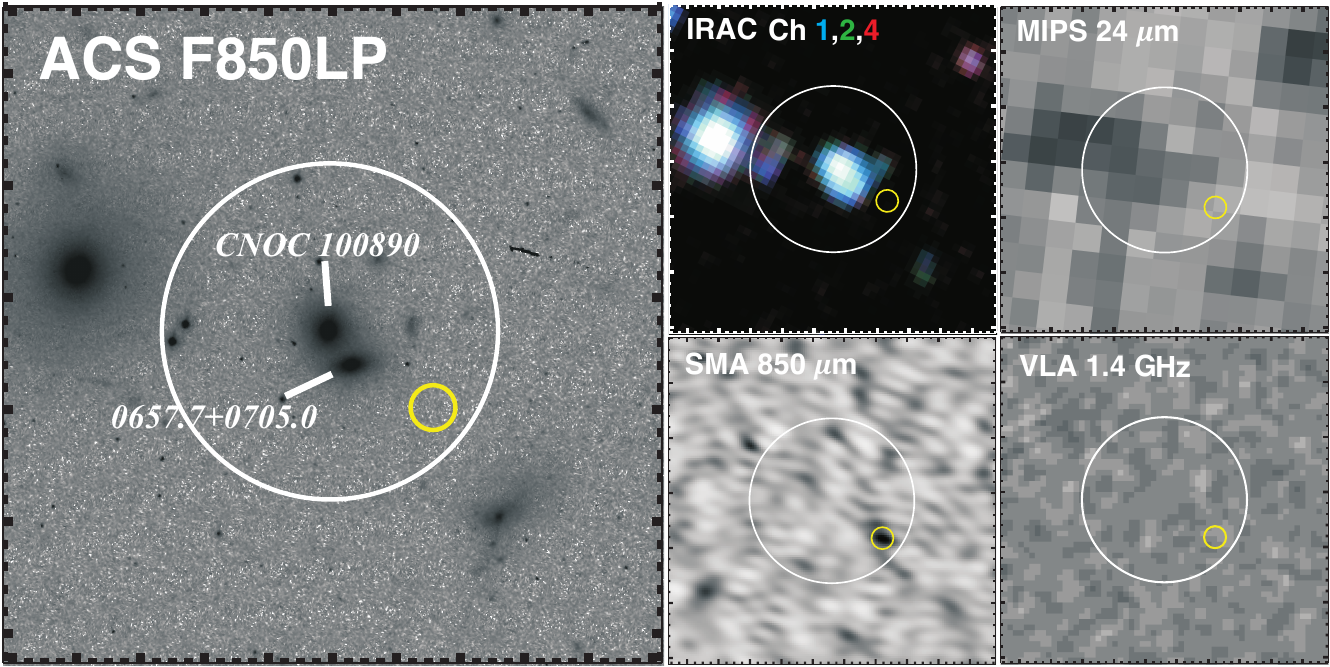}
 	\end{center}
       \caption{ Multiwavelength images of A2390-3 from 
\citet{Chen:2011p11605}. North is up and east is to the left. The size of each image is 30$\arcsec \times 30\arcsec$ . The big white circle in each image is the SCUBA beam size (15 $\arcsec \times 15\arcsec$ ). The SMA position of A2390-3 is labeled in each image with a yellow circle. The gray-scale images have inverse scales. Three of the IRAC channels (1, 2, and 4), corresponding to 3.6, 4.5 and 8.0 $\mu$m, are presented in the 
combined IRAC image with the color codes labeled.}
     \label{a23903}
\end{figure}

\section{Discussion}
The coarse resolution of the 850~$\mu$m maps from single-dish survey has always been a major issue on the studies of the SMGs. It not only leads to confusion limits, it also makes locating the true counterparts of the SMGs at other wavelengths and deciding the accurate amplification for those strongly lensed SMGs difficult: Using high resolution interferometric imaging with the SMA to resolve blank-field SCUBA detected sources, \citet{Wang:2011p9293} found that the multiplicity appears to be common and the selection techniques from other wavelengths are usually inadequate. A similar study by \citet{Chen:2011p11605} on several strongly lensed faint SMGs shows that the large uncertainty on the source amplifications caused by the poor resolution can lead to severe misinterpretations about their nature.   

What is more important is that, like A2390-5, the only faint SMG detected in \citet{Chen:2011p11605} shows no counterpart in other wavelengths either (Figure \ref{a23903}), which leads to the fact that so far all SMA-observed faint SMGs from the non-preselected sample are not traceable. This happens much more frequently than it is expected from bright SMGs (\citealt{Younger:2007p6982, Younger:2009p9502}), meaning that we are detecting an unexpectedly large number of high redshift dusty galaxies. This could be of considerable importance given that most studies on faint SMGs use selection techniques from other wavelengths and their results could be misleading if many faint SMGs are not traceable in this way. We therefore emphasize the importance of direct submillimeter high-resolution studies of the faint SMGs, which could be the dominant population of the high-z star formation.

\bibliography{bib}
\end{document}